\documentclass[final, 3p, authoryear]{elsarticle}
  \journal{Journal of Theoretical Biology}
\usepackage[nodots]{numcompress}
\usepackage{amsmath,amssymb,amsthm,amsfonts}
\usepackage[]{graphicx}
\usepackage{setspace} 
\usepackage{lineno} 
\linenumbers

\def\citeapos#1{\citeauthor{#1}'s (\citeyear{#1})}
\def\tableline{\vskip .1in \hrule height 0.6pt \vskip 0.1in}
\usepackage{hyperref}

\begin{document}

\begin{frontmatter}
\title{Environ centrality reveals the tendency of indirect effects to
homogenize the functional importance of species in ecosystems\tnoteref{t1}} 

\tnotetext[t1]{SLF contributed methods, analyzed data, co-wrote the
  paper.  SRB conceived the research design, contributed methods,
  analyzed data, and co-wrote the paper.}

\author[bio,math]{Sarah L. Fann} 

\author[bio,cms]{Stuart R. Borrett \corref{cor1}}
\ead{borretts@uncw.edu}

\cortext[cor1]{Corresponding author. Tel. 910.962.2411; fax: 910.962.4066}

\address[bio]{Department of Biology \& Marine Biology, University of
  North Carolina Wilmington, 601 S.\ College Rd., Wilmington, 28403
  NC, USA} 
\address[math]{Department of Mathematics and Statistics, University
  North Carolina Wilmington, 601 S.\ College Rd., Wilmington, 28403
  NC, USA}
\address[cms]{Center for Marine Science, University of North Carolina
  Wilmington, 601 S.\ College Rd., Wilmington, 28403
  NC, USA} 

\begin{abstract}
  Ecologists and conservation biologists need to identify the relative
  importance of species to make sound management decisions and
  effectively allocate scarce resources. We introduce a new method,
  termed environ centrality, to determine the relative importance of a
  species in an ecosystem network with respect to ecosystem
  energy--matter exchange.  We demonstrate the uniqueness of environ
  centrality by comparing it to other common centrality metrics and
  then show its ecological significance. Specifically, we tested two
  hypotheses on a set of 50 empirically-based ecosystem network
  models.  The first concerned the distribution of centrality in the
  community.  We hypothesized that the functional importance of
  species would tend to be concentrated into a few dominant species
  followed by a group of species with lower, more even importance as
  is often seen in dominance--diversity curves.  Second, we tested the
  systems ecology hypothesis that indirect relationships homogenize
  the functional importance of species in ecosystems. Our results
  support both hypotheses and highlight the importance of detritus and
  nutrient recyclers such as fungi and bacteria in generating the
  energy--matter flow in ecosystems. Our homogenization results
  suggest that indirect effects are important in part because they
  tend to even the importance of species in ecosystems. A core
  contribution of this work is that it creates a formal, mathematical
  method to quantify the importance species play in generating
  ecosystem activity by integrating direct, indirect, and boundary
  effects in ecological systems.
\end{abstract}


\end{frontmatter}
\begin{spacing}{1}
\section{Introduction}
%

Identifying the functional importance of species is a fundamental step
in describing community and ecosystem structure and function.  It is
essential for ecologists, conservation biologists, and resource
managers to understand species relative importance so that they can
make informed management decisions and effectively allocate limited
conservation resources.  Rank-abundance curves \citep{whittaker65} are
a classic tool used to describe the biodiversity and relative
importance of species in a community.  This approach generally assumes
that the importance of the species is proportional to its abundance or
biomass.  Alternative techniques include using functional measures
like productivity in dominance--diversity curves \citep{whittaker65},
and more recent concepts such as network role equivalence
\citep{luczkovich03}, keystone species \citep{power96}, ecosystem
engineers \citep{lawton94, jones94}, and more generally the concept of
foundational species \citep{dayton72, ellison2005loss}. However,
quantifying the relative functional importance of species embedded in
their communities remains a challenging problem.  In this paper, we
introduce a new approach to quantifying the relative functional
importance of species in ecosystem networks termed \emph{environ
  centrality}. While species importance can be defined in a variety of
ways, here we specifically focus on their importance for
energy--matter distribution in communities.
While we focus on ecosystems, the methods are generally applicable to
any complex system of conservative fluxes modeled as a network.

One reason that the fundamental step of quantifying species functional
importance has not been fully addressed is that indirect interactions
complicate the assessment.  Organisms are embedded in an intricate web
of interactions and it is this tangled network of relationships that
enables indirect influences to become significant components of
ecological and evolutionary interactions \citep[e.g.][]{patten84,
  wootton94_review, whitham2006framework, bascompte07, estrada08,
  montoya09, keith2010genetic}.  For example, \citet{poulin10} showed
how invasive species can increase the prevalence and severity of
disease in a community through trait-mediated indirect
effects. Trophic cascades are one type of trophic indirect effect that
can have large and unexpected impacts on ecosystems
\citep{carpenter85, mooney10}. For example, \citet{berger08} found
that where wolf populations were extripated, mesopredator populations
like coyotes were released from predator control.  This then resulted in a
four fold decrease in the survival rate of pronghorn fawns, a
coyote prey item.
Thus, understanding the ecological and functional importance of
species in their ecosystem requires understanding the full environment
of direct and indirect influences.

Network models of ecosystems let ecologists quantify species importance and
are a key tool for determining the importance of indirect
influences that are distributed through these types of interactions
\citep[e.g.][]{bondavalli99, dame06, belgrano05}. In these models,
network \emph{nodes} represent species, groups of species, or abiotic
resource pools.  Although the nodes may be a species complex or
non-living compartment of energy--matter, we generically refer to them
as species throughout this paper for simplicity.  Network \emph{links}
represent the flow of energy--matter from one node to another. These
energy--matter flows can be created by several ecological processes
including feeding, death, and excretion.  This representation of
complex ecosystem interactions lets ecologists apply mathematical and
computational tools to learn more about the structure and function of
ecosystems. These ecosystem networks can be represented by weighted
and directed graphs so that a link not only implies a relationship
between two nodes but it also indicates \textit{how much} (weighted)
energy--matter and \textit{from whom to whom} (directed) the
energy--matter is flowing.

Ecological Network Analyses (ENA) are a collection of quantitative
methods that systematically asses information from a full, complex
network description. There are several specific implementations of
this concept, such as Ecopath \citep{christensen04}, Network Environ
Analysis \citep{fath99_review,fath06}, EcoNet \citep{schramski11}, and
NETWRK \citep{ulanowicz86}. ENA's throughflow analysis mathematically
partitions ecosystem energy--matter flow across boundary, direct, and
indirect paths in the network \citep{fath99_review, fath06,
  schramski11}.  This approach has been applied to analyze the
structure and function of ecosystems.  For example, \citet{finn80}
used the technique to characterize mineral cycling in the Hubbard
Brook ecosystem.  More recently, \citet{borrett06_neuse} and
\citet{schramski06} investigated nitrogen cycling in the Neuse River
Estuary, and \citet{zhang10} investigated the urban water metabolic
system of Beijing.  ENA lets scientist study the efferent and afferent
holistic environments of species within a system boundary, which
Patten termed environs \citep{patten78}.

Based on the development and application of ENA, systems ecologists
have hypothesized that indirect energy--matter flows tend to dominate
direct flows in ecosystems \citep{higashi89, jorgensen07_newecology}.
\citet{fath04_cyber} found support for this hypothesis in large
ecosystem models generated from a community assembly type algorithm,
\citet{borrett06_neuse} found supporting evidence for the hypothesis
in 16 seasonal nitrogen cycling models for the Neuse River estuary,
and \citet{salas11_did} found general support for the hypothesis in 50
empirically-based trophic ecosystem models.  A consequence of these
large indirect energy--matter flows is that resources tend to be more
evenly distributed in the systems \citep{fath99_homo,
  borrett10_hmg}. Given this tendency, we hypothesized that indirect
effects tend to homogenize the relative importance of the species,
decreasing the relative influence any single species has on ecosystem
functioning.

This paper has three main objectives.  First, we introduce a new
metric of functional importance based on the throughflow analysis of
ENA and centrality concepts from social science.  We contrast this
measure with other existing centrality indexes to demonstrate its
utility and uniqueness.  Second, we use this measure to characterize
the relative importance of species in 50 trophic ecosystem models.
Third, we test two hypotheses regarding the functional organization of
ecosystems.  Based on previous community analysis, we first
hypothesized that there will tend to be a ``concentration of
dominants'' or functionally more important species and a ``long tail''
of species with a lower and more similar importance.  We expected that
the ecosystem dominants would be compartments like detritus,
particulate organic matter, and bacteria because of their critical
role in recirculating energy--matter and connecting the green and
brown food webs \citep{wilkinson06, jordan07}. We further tested the
hypothesis that indirect effects tend to homogenize the relative importance
of species.

This paper is organized as follows.  In the next section we introduce
the centrality concept and describe its use and diversity.  We then
describe the ENA throughflow analysis that we use and introduce the
environ centrality measure. Section 4 describes our analytical
approach to testing our hypotheses and Section 5 describes our
results.  We conclude the paper by putting these results into a
broader context.

\section{Centrality} 
Centrality is a concept used by scientists studying complex systems
that was initially developed by social scientists to quantify the
importance of individuals in network models \citep{wasserman94,
  borgatti05}. Metrics based on this concept indicate how a node's
position in the network contributes to its structural or functional
importance.  There are multiple centrality metrics with varying
characteristics \citep{wasserman94, borgatti06}, but many tend to be
well correlated \citep{valente08, newman06}.  Despite this
correlation, differences between the measures can be useful.  For
example, degree centrality is based on the number of network edges
directly touching a node and describes the local or proximate
importance of the node.  In contrast, eigenvector centrality describes
the stable distribution of pathways (i.e. at long pathways) passing
through the nodes, which provides a more global or whole network
understanding of the nodes' relative importance. These different
measures have been well described in the literature, so here we focus
on the ecological relevance of selected centrality measures.

There are several approaches to quantitatively describing ecosystem
networks \citep{ulanowicz86, bersier02}. Several metrics commonly used
to describe food webs such as link density and connectance
\citep{dunne02} are what \citet{wasserman94} call group level
indicators of centrality.  Link density is the average node degree in
the network.  Connectance is link density proportional to the size of
the network.
Degree centrality, and thus link density and connectance, is a
\emph{local} centrality metric because it only considers the
immediately adjacent neighborhood of connected nodes.  Local
centrality metrics may be a useful starting point for describing a
food web, but they may not be the best descriptors of the organism's
importance in energy--matter transference because they assume the
influence is restricted to a local neighborhood.  Thus, they neglect
the important indirect influences.  From a trophic point of view,
indirect effects become important in part because energy--matter is
passed from one organism to another and may ultimately reenter the
same organism through nutrient cycling, creating a well connected web
of interactions \citep{allesina05_scc, borrett07_jtb}.

Centrality metrics that incorporate indirect influences include
betweenness centrality \citep{freeman79}, eigenvector centrality
\citep{bonacich72, borgatti05, estrada08, allesina09}, and weighted
topological importance \citep{bauer10, jordan03}.  These centrality
metrics are more appropriate for many ecological applications than
those that only consider direct influences, but still have important
limitations.  Betweenness centrality considers the possible gatekeeper
role that some nodes may play in the transfer of information, energy,
and matter from two more distinct groups.  While this metric can be
useful, its focused on a particular type of bridging importance
\citep{wasserman94} and was not intended to provide a more general
metric of the importance of a node.  In trophic or biogeochemical
network models, eigenvector centrality effectively quantifies the
equilibrium distribution of energy--matter flowing through the nodes,
thus considering the direct and indirect interactions. However, there
are two potential problems with this measure for ecological networks.
First, it ignores the contribution of the initial transient effects
that maybe important in some contexts, especially systems in which the
strength of interaction decreases rapidly with path length like
ecosystems \citep{borrett10_idd}.  Second, it only considers the
dominant eigenvector, which depending on the structure of the network
may not be an adequate approximation of the transfer dynamics
\citep{borgatti06}.  Weighted topological importance quantifies the
effect a species has on others in a community, which is great for
understanding competition but does not provide information on how
species contribute to global network properties such as total
energy--matter throughflow. To address these limitations, we introduce
\emph{environ centrality}, which uses weighted information, integrates
direct and indirect effects, describes how species contribute to
global network measures, and captures the transient dynamics as well
as the equilibrium effects.

\section{Ecological Network Analysis} 


\subsection{Throughflow Decomposition}
As comprehensive summaries of ENA methodology exist
\citep{ulanowicz86, fath06, schramski11}, here we focus on
the components of ENA necessary to calculate the centrality of species in
ecosystem network models.  We used the Network Environ Analysis Matlab$^{\circledR}$
function \citep{fath06} to implement these analyses. 

Let a network model with $n$ nodes be represented by a matrix,
$\mathbf{F}_{n\times n} = (f_{ij})$, which defines the quantity of
energy--matter being transferred from node $j$ to node $i$. The
structural component of $\mathbf{F}$ is the adjacency matrix,
$\mathbf{A} = (a_{ij})$, in which a 1 indicates a direct connection
from $j$ to $i$ and 0 indicates none. Energy--matter entering the
system at node $i$ is denoted by $z_{i}$, whereas energy--matter
leaving the system at node $j$ is $y_{j}$.

ENA uses this system data to determine several ecosystem
properties. First, the total flow through a node is defined as
throughflow, which can be input $T_{i} = z_{i} + \sum_{j=1}^n f_{ij}$
or output $T_{j} = \sum_{i=1}^n f_{ij} + y_{j}$ oriented. ENA
typically assumes that the networks are at steady state
(e.g., mass-balanced) so $T_{i} = T_{j}$. Second, we calculate the
direct flow intensity matrix $\mathbf{G}$. We focus on the output
oriented direct flow intensity matrix, which is calculated as
$\mathbf{G} = (g_{ij}) = f_{ij}/{T_j}$.  Elements of $\mathbf{G}$ are
unitless and represent the direct flow intensity from node $j$ to node
$i$. Third, raising $\mathbf{G}$ to a power $m$, gives the flow
intensity between two compartments over all paths of length $m$.  The
sum of flow intensities over all pathways in a network is defined as
the integral flow matrix $\mathbf{N}$, which is
\begin{linenomath}
\begin{equation}
  \mathbf{N}\equiv \sum_{m=0}^\infty{\mathbf{G}^m} =  \underbrace{\mathbf{G}^0}_{\textrm{Boundary}} + \underbrace{\mathbf{G}^1}_{\textrm{Direct}} + \underbrace{\mathbf{G}^2+\ldots+\mathbf{G}^m+\ldots}_{\textrm{Indirect}}.  \label{eq:N}   
\end{equation}
\end{linenomath}
$\mathbf{N}$ quantifies the intensity of output oriented throughflow
from $j$ to $i$ over all pathways in the network.
Since ecological systems are thermodynamically open, the
energy--matter must dissipate as path length increases causing the sum
of flow intensities over all pathways to converge.  Thus, we can use
the identity $(\mathbf{I}-\mathbf{G})^{-1}$, where $\mathbf{I} =
\mathbf{G}^0$ is the identity matrix, to find the exact values of
$\mathbf{N}$.

Multiplying the integral matrix $\mathbf{N}$ by the input vector
$\mathbf{z}$ returns the output oriented node throughflow $\mathbf{T}
= \mathbf{Nz}$.  This equation ensures that equation (\ref{eq:N}) is a
true partition of flow and that the flow elements are not double
counted. Total System Throughflow ($TST = \sum_{j=1}^n T_{j}$) is the
sum of node throughflows and is a global measure of the total network
activity or size of the system. Thus, the integral matrix shows how
$TST$ is generated by species in an ecological network and
incorporates energy--matter flux over all indirect pathways. Environ
centrality is calculated from the integral flow matrix and quantifies
the relative importance of species in creating total system activity.
   
\subsection{Environ Centrality}
Environ centrality is a measure based on the ENA output-oriented
integral flow matrix, $\mathbf{N}$.  We introduce three related
environ centrality measures:  input ($EC^{in}$), output ($EC^{out}$),
and an average of the two ($AEC$). These are calculated as follows:
\begin{linenomath}
\begin{align}
\notag EC^{in} &= \frac{\sum_{j=1}^n n_{ij}}{\sum_{i=1}^n\sum_{j=1}^n  n_{ij}};  \\ 
\notag EC^{out} &= \frac{\sum_{i=1}^n n_{ij}}{\sum_{i=1}^n\sum_{j=1}^n n_{ij}}; \textrm{ and} \\ 
AEC &= \frac{EC_{i}^{in} + EC_{j}^{out}}{2}. \label{eq:EC}
\end{align}
\end{linenomath}
$EC^{in}$ and $EC^{out}$ indicate the relative importance of species in
generating ecosystem activity from an input or output direction,
respectively. $AEC$ determines the relative importance of species in
ecosystem models by averaging the input and output importance. Thus, $AEC$
quantifies the relative contribution of a node to the energy--matter
exchange within a system. It is a global centrality measure of the
functional importance of species because it incorporates direct and
indirect pathways, weighting pathways by the amount of energy--matter
flux intensity passing through a node.  Further, it captures the
transient dynamics as well as the equilibrium dynamics and describes
how a species contributes to total system activity. Fig.~\ref{fig:rank}
shows an example rank-order distribution of $AEC$ for the Georges Bank
ecosystem \citep{link08}.

ENA is an Input--Output analysis in which it is possible to calculate
output oriented and input oriented integral flow intensities
\citep{fath06, schramski11}.  Thus, it is possible to calculate
the $EC^{in}$, $EC^{out}$, and $AEC$ for both of these
orientations. The interpretation of output and input oriented integral
flow intensities differ.  The output orientation looks forward in time
and follows the direction of the network arrows.  The input orientation
looks backward through the network, against the arrow directions, to
see where the flow has come from.  Both orientations can be
analytically useful independently and when jointly considered
\citep{ulanowicz90,schramski06}.
As the purpose of this paper is to introduce environ centrality and
illustrate its ecological significance, only the output oriented case
is analyzed for simplicity.  We expect the input oriented results to
be similar, although the interpretation will differ.
	
\section{Experimental Design}	
In this section we describe our experimental design to test our
hypotheses.  We first characterize the set of 50 empirically based
ecosystem models we analyzed.  We then explain our evaluation of the
utility and uniqueness of $EC$.  Finally, we describe our approach for
determining species dominance and the centrality homogenization in 50
ecosystem networks.

\subsection{Network Models}
We calculated $AEC$ for 50 empirically-based network models of
ecosystem energy--matter flows (Table~\ref{tab:models}).
\citet{borrett10_hmg} first assembled this data set to test the
distinct systems ecology resource homogenization hypothesis. The
collection is available from
\href{http://people.uncw.edu/borretts/research.html}{http://people.uncw.edu/borretts/research.html}.
Models were selected that spanned a range of sizes from smaller,
highly aggregated systems comprising 4 nodes to larger, less
aggregated systems comprising 125 nodes. Each network model traces
trophic and non-trophic energy--matter transactions such as feeding,
excretion, mortality, immigration and emigration and together
represent 35 distinct ecosystems.  To avoid a selection bias, all
models discovered in the literature that were created to model a
specific ecosystem and that included empirical data were included in
the data set \citep[see][for more details]{salas11_did}. To meet the
ENA steady-state assumption, we balanced 15 of the 50 models using the
AVG2 algorithm, which \citet{allesina03} demonstrated minimized
balancing bias.

\subsection{AEC Sensitivity and Uniqueness}
Given the environ centrality metric, our first step was to evaluate
its sensitivity and demonstrate its novelty.  To illustrate its
sensitivity, we compared $AEC$ values for four realizations of a
hypothetical ecosystem flow network (Fig.~\ref{fig:toy}).  The four
realizations vary in their distribution of flow amongst the five
internal direct pathways.  This is evident in the direct flow
intensity matrices, $\mathbf{G}$, and the subsequent integral flow
matrices, $\mathbf{N}$.  The direct flow intensity values in the
second realization are 10\% of those in the first, which is a
universal extensive change.  The distribution of flow intensities is
the same, but in the second realization the total magnitude is less.
Realizations three and four have equal total flow intensities, but
differ in how it is distributed.  Thus, this is an example of an
intensive change.  Together, these realizations demonstrate $AEC$'s
ability to capture both intensive and extensive network changes. In
addition to $AEC$, we also examined the average eigenvector centrality
$AEVC$ (Table~\ref{tab:cent}) since these were likely to be the most
similar centrality measures.

\begin{figure*}[!ht]
  \centering
  \includegraphics[scale=0.75]{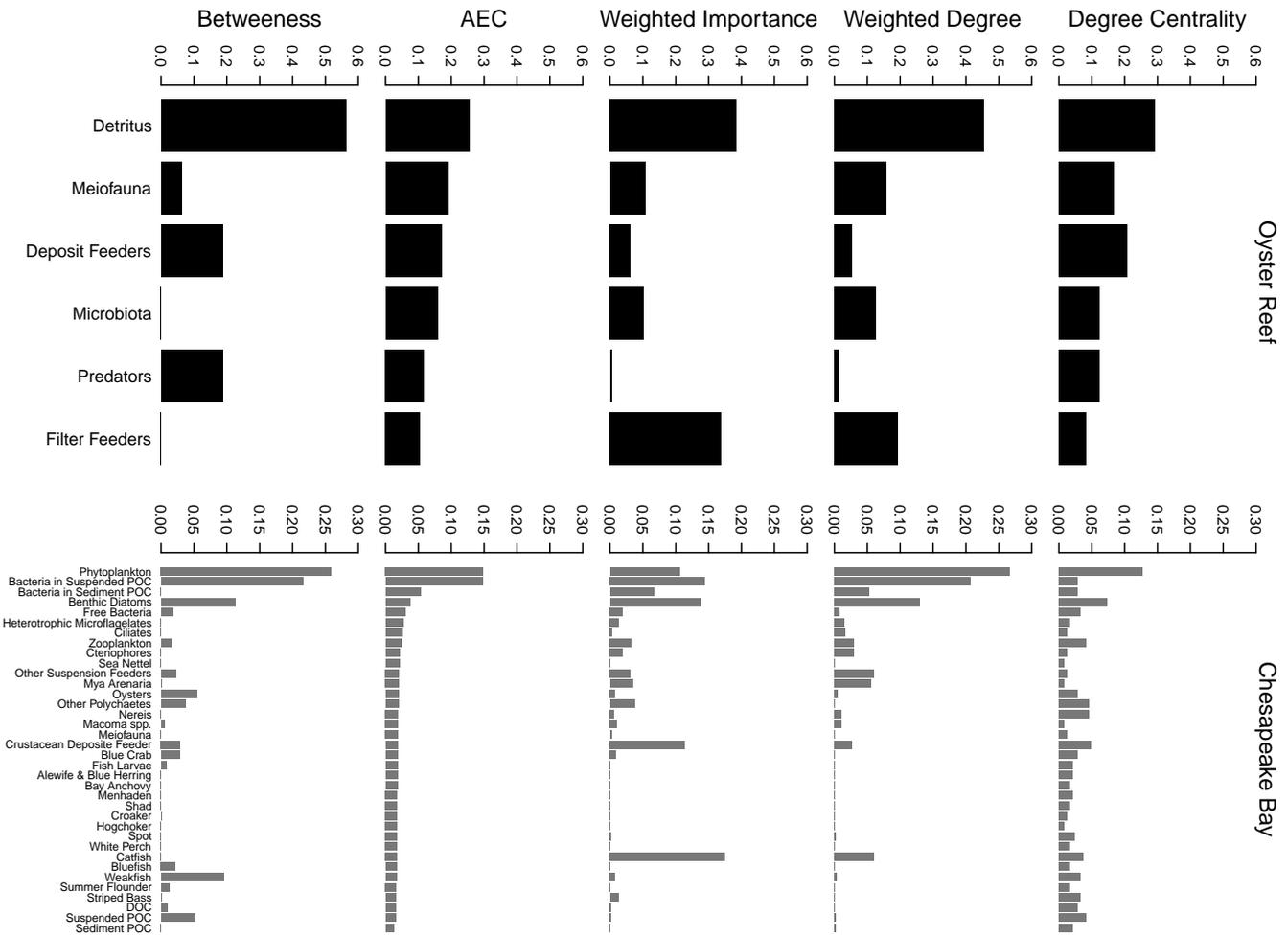}
  \caption{$DC$, $WD$, $WI$, $AEC$, and $BC$ distributions for Oyster
    Reef and Chesapeake Bay network models.  Model nodes are ordered
    according to $AEC$ rank.}
  \label{fig:comp}
\end{figure*}

To show $AEC$'s uniqueness, we compared its results in two ecosystem
models to five commonly used centrality metrics: degree centrality
($DC$), weighted degree centrality ($WD$), weighted topological
importance ($WI$), and betweenness centrality ($BC$).  The Oyster Reef
and Chesapeake Bay models were selected as case studies to demonstrate
the distributions of these five centrality metrics in empirical
network models (Fig.~\ref{fig:comp}). We calculated all metrics using
common formulations (Table~\ref{tab:cent}) and then normalized each
set of scores by its sum \citep[{\emph{sensu}}][]{ruhnau00}, creating
proportions to facilitate comparisons among the five metrics. In
addition to the comparisons in Fig.~\ref{fig:comp}, we compared $AEC$ and
$AEVC$ for four network models; the Oyster Reef, Chesapeak Bay,
Bothnian Bay, and Northern Benguela.  Finally, we calculated the
Spearman rank correlation of the $AEC$ and $AEVC$ in the 50
empirically based ecosystem models listed in Table~\ref{tab:models}.
We focus on these two centrality measures because they are most likely
to be similar because they are both global pathway centralities.

\subsection{Dominance and Evenness}
To address our hypothesis regarding the relative importance of species in
ecosystems, we used the coefficient of variation ($CV$) to
characterize the evenness of the $AEC$ distribution and to identify
dominant species in 50 ecosystem models. Although there are multiple
methods for quantifying variation, we chose $CV$ because it is a
non-dimensional measure normalized to the mean values.  This lets us
compare the relative variation across systems, even if they differ in
the number of species or $TST$.  In addition, $CV$ is sensitive to
distribution skew \citep{fraterrigo08}, which we use to our advantage
to identify dominant species.

We considered ecosystems with a $CV$ less than unity (1) to be low
variance.  To identify the dominant species, we ranked species
according to their $AEC$ scores and calculated the $CV$ for the entire
community.  We then progressively removed the highest ranking species
until the $CV$ of the remaining community $AEC$ was less than or equal
to $0.5$.  A $CV$ value of 0.5 is low enough to identify dominant
species without being so low as to claim all species are dominant. It
also identifies which species are responsible for the highest
proportion of the variation in $EC$. All species removed before
reaching the $CV$ threshold are classified as dominant species with
respect to the ecosystem flux of energy--matter. Thus, this approach
determines both the evenness of non-dominant species, and identifies
the dominant species.

\subsection{Homogenization and Indirect Effects}
Our final hypothesis was that indirect flows would homogenize the
relative importance of species in generating $TST$.  We addressed this
hypothesis by comparing the relative importance of species in the
direct flows to those of the integral flows.  Recall that
equation~\ref{eq:N} calculates the integral flow intensity matrix
$\mathbf{N}$ that combines flow intensity over all boundary
($\mathbf{G}^0$), direct ($\mathbf{G}^1$), and indirect pathways
($\sum_{m=2}^\infty\mathbf{G}^m$).  To find the relative importance of
species from the perspective of the direct flows alone, we substituted
$\mathbf{G}$ for $\mathbf{N}$ in equation~\ref{eq:EC} to calculate
\emph{direct} average environ centrality ($AEC^{direct}$). We then
compared $AEC^{direct}$ to the integral average environ centrality
($AEC$) to test the homogenizing effect of indirect relationships on
$AEC$.

As the integral matrix includes boundary, direct, and indirect flow
intensities, it is possible that observed homogenization could be
caused by the boundary input as well as the indirect flows. To isolate
the effect of the indirect flows, we also compared the $AEC^{direct}$
to the $AEC$ calculated on $\mathbf{N}-\mathbf{G}^0$ instead of
$\mathbf{N}$.
	
Homogenization of species importance was quantified by comparing the
$CV$ of $AEC^{direct}$ and the $CV$ of $AEC$.  We created a ratio of
the two, $CV(AEC^{direct})/CV(AEC)$, such that when the ratio is
greater than unity (1) it indicates that the $AEC$ is more evenly
distributed than $AEC^{direct}$. Ratio values greater than unity
indicate that indirect effects homogenize the importance of species in
generating ecosystem activity.

\section{Results}  

\subsection{EC Sensitivity and Uniqueness}
\begin{figure*}[!t]
  \centering
  \caption{Example of average environ centralities ($AEC$) in four
    realizations of a hypothetical ecosystem model.  (A) Model
    structure, (B) four mass balanced realizations of the direct flow
    intensity matrices, $\mathbf{G}$, for the model, (C) corresponding
    integral flow matrices, $\mathbf{N}$, (D) $AEC$ distributions, and
    (E) average eigenvector centralities $AEVC$.  Notice that flow is
    from column $j$ to row $i$ in the matrices. When column sums of
    $\mathbf{G}$ do not equal 1, the remaining portion of flow is lost
    as exports or respiration.}
  \includegraphics[scale=0.7]{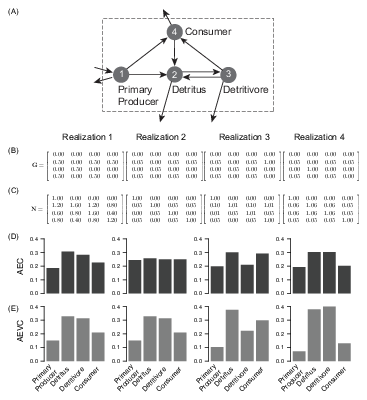}   
  \label{fig:toy}

\end{figure*}

To establish the sensitivity of $AEC$ to both intensive and extensive
changes, we applied it to four realizations of the hypothetical
ecosystem model shown in Fig.~\ref{fig:toy}.  The $AEC$ distributions
for the realizations are clearly different (Fig.~\ref{fig:toy}D).  In
the first realization, the detritus box was more important and
the importance of the primary producer was diminished.  In the second
realization in which the direct flow intensities were reduced by 90\%,
the $AEC$ values are much more similar.  This is because the drop in
flow intensities is then transmitted through the longer pathways,
effectively discounting them. However, there is no difference between
the $AEVC$ centrality distributions between realization one and two,
demonstrating that eigenvector centrality is not sensitive to
this extensive change.

Realizations three and four maintained the total magnitude of the
network but have different distributions, which is an intensive
change. The $AEC$ and $AEVC$ distributions between realization three
and four are different, demonstrating that both metrics are sensitive
to these intensive changes.  In the third realization, one flux to the
consumer compartment increased which caused it to become the second
most central compartment.  The detritus compartment also responded to
the change, increasing its centrality from 0.26 to 0.30. For the
fourth realization, the flux from detritivore to detritus increased,
causing detritus to maintain its centrality from the previous
realization at 0.30 and detritivores to increase its centrality to
0.30 as well. These four cases show that $AEC$ is sensitive enough
to capture important differences between the model flow realizations,
and specifically demonstrates its ability to capture both intensive
and extensive changes to direct energy--matter flows.

   
The comparison of the four centrality metrics to $AEC$ in the Oyster
reef and Chesapeake Bay ecosystem models (Fig.~\ref{fig:comp}) shows
the different structural and functional importance of the species
within each network and illustrates the unique perspective of $AEC$.
$DC$ highlights which species were most connected within the network
topology (unweighted), yet was not always a predictor of which species
would be most important in generating $TST$. There were several
disparities between $DC$ and $AEC$ which show that being the most
connected does not necessarily result in being a key species in
generating $TST$. For example, the suspended particulate nutrient node
from the Chesapeake Bay model has a greatly reduced ranking in $AEC$
despite having the second largest $DC$.  $WD$ had similar mismatches
with $AEC$, and demonstrates that high or low magnitudes of direct
flows are not always indicative of overall importance in generating
ecosystem activity. In both models, $WI$ consistently placed primary
producers and herbivores (i.e. filter feeders, phytoplankton, and
zooplankton) as species with the highest effect on others, yet these
species ranked low in $AEC$. The $BC$ distribution was the most
distinct centrality metric.

\begin{figure*}[!t]
  \centering
  \includegraphics[scale=0.7]{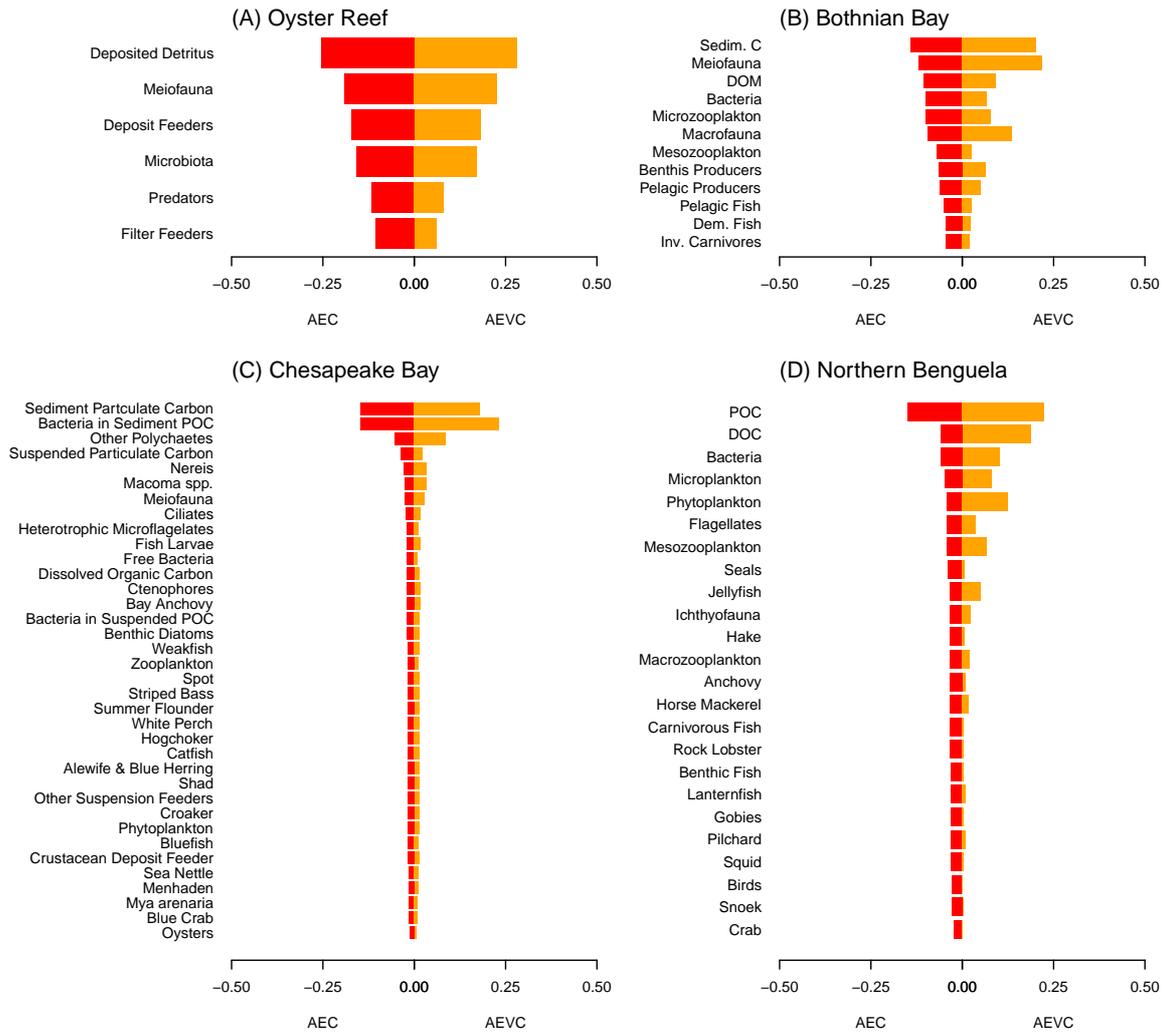}
  \caption{Comparison of the average environ centrality ($AEC$) and average eigenvector
    centralities ($AEVC$) for the (A) Oyster Reef (B) Bothnian Bay (C)
    Chesapeake Bay and (D) Northern Benguela ecosystem models.  The
    nodes have been ordered in ascending order of the average environ
    centrality to facilitate comparison.  The $AEC$ distributions were
    multiplied by -1 for plotting convenience.} \label{fig:comp2}
\end{figure*}

Average eigenvector centrality should be the most similar to average
environ centrality because they are both global centrality measures
related to pathways or energy--matter flow; however, we see from
Fig.~\ref{fig:toy} that they can differ. Fig.~\ref{fig:comp2} further
illustrates differences between $AEVC$ and $AEC$.  It shows both types
of centralities for four cases: the Oyster Reef and Chesapeake Bay
models used in Fig.~\ref{fig:comp} and the Bothnian Bay and Northern
Benguela ecosystem models.  The centralities indicate a difference of
the rank-order importance of the nodes for all models except the
Oyster Reef.  For example, $AEC$ suggests that Bacteria, DOM,
Mesozooplankton, and Microzooplankton have a larger functional
importance in the Bothnian Bay ecosystem than is suggested by the
eigenvector centrality (Fig.~\ref{fig:comp2}B).  In contrast,
eigenvector centrality indicates a very dominant importance of
Meiofauna, Sediment carbon, and Macrofauna.  The two centralities also
show a different rank importance of the nodes in the Northern Benguela
ecosystem model. From the $AEC$ perspective, POC has a dominant
importance, while most of the other species have very similar
centralities.  Variation in $AEVC$ is larger in all models, including
Oyster Reef where the rank importance are the same.  One interesting
difference is the diminished importance of Seals suggested by $AEVC$
for Fig.~\ref{fig:comp2}D). Additionally, $AEVC$ ranks bacteria in
sediment POC as more important than Sediment Particulate Carbon in
Fig.~\ref{fig:comp2}C while the $AEC$ values are nearly the same.

To generalize our analysis, we calculated the Spearman rank
correlation between $AEC$ and $AEVC$ for the 50 ecosystem network
models (Table~\ref{tab:models}).  These correlations ranged from 0.2
to 1 and had a median value of 0.92.  As we expected, this suggests
that these two centrality measures are typically, but not always,
similar.  Our case studies in Fig.~\ref{fig:comp2} highlight how these
small correlation differences can be ecologically meaningful.  

\subsection{Dominance and Evenness}
\begin{figure*}[!t]
  \centering
  \includegraphics[scale=1]{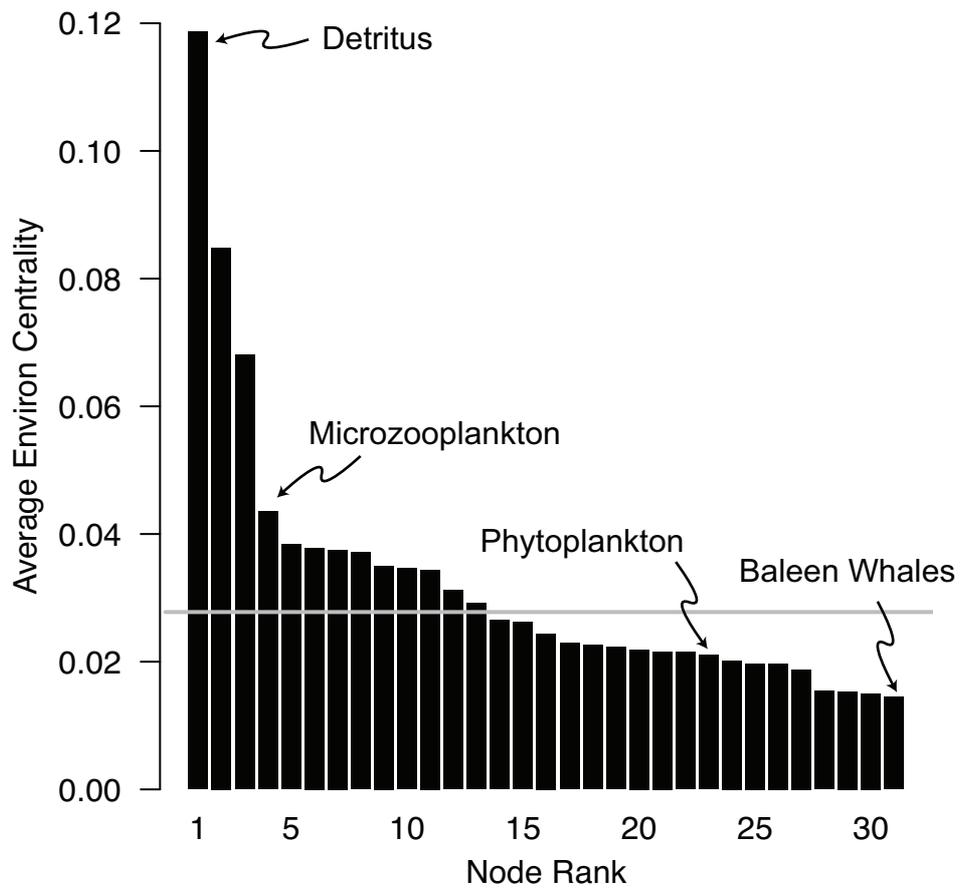}
  \caption{Rank $AEC$ curve of the Georges Bank Ecosystem network
    model. The grey line indicates the potential even
    distribution. Notice that the area under the curve sums to one.}
  \label{fig:rank}
\end{figure*}

Rank $AEC$ curves provide a tool to visualize the relative importance of
species (Fig.~\ref{fig:rank}).  The $AEC$ distribution of
the Georges Bank illustrates the tendency for a few nodes to have
high $AEC$ with most having relatively lower and even $AEC$
values. $AEC$ variation in the ecosystem models we analyzed was
generally low (Table~\ref{tab:dom}).  Twelve of the models have a
$CV(AEC)$ less than $0.5$ and 30 of the models have a $CV(AEC)$
between $0.5$ and unity.  However, eight of the models exhibit more
variability with CV values larger than unity.  In the models with a CV
larger than $0.5$, no more than three dominant species had to be
removed before the $CV$ of the remaining species fell below $0.5$
(Table~\ref{tab:dom}).  The median number of dominants was two. As
expected, detritus and detrital recyclers are predominantly responsible
for generating $TST$ in most ecosystems, with water flagellates being
the only non-detrital or bacterial species to rank as dominant $AEC$
contributors.


The second hypothesis was that sub-dominant species would have a more
even distribution of importance. In Fig.~\ref{fig:rank}, the first two
species were considered dominant species and the $CV$ score for the
whole ecosystem was 0.68.  However, once the dominants were removed
the $CV$ score of the remaining species was $0.41$ suggesting less
variation in the importance of the remaining species. Table~\ref{tab:dom}
identifies the $CV$ value of the entire community, as well as the
community without the dominant species for all 50 ecosystem
models. After the dominant species were removed, the $AEC$ scores were
relatively even, with an average $CV$ score of $0.38$. This indicates
that the standard deviation was only 38\% the magnitude of the mean.
All the networks support our hypothesis that the importance of sub-dominant
species would be relatively even.

\subsection{Homogenization and Indirect Effects}
\begin{figure*}[!t]
  \centering
  \includegraphics[scale=1]{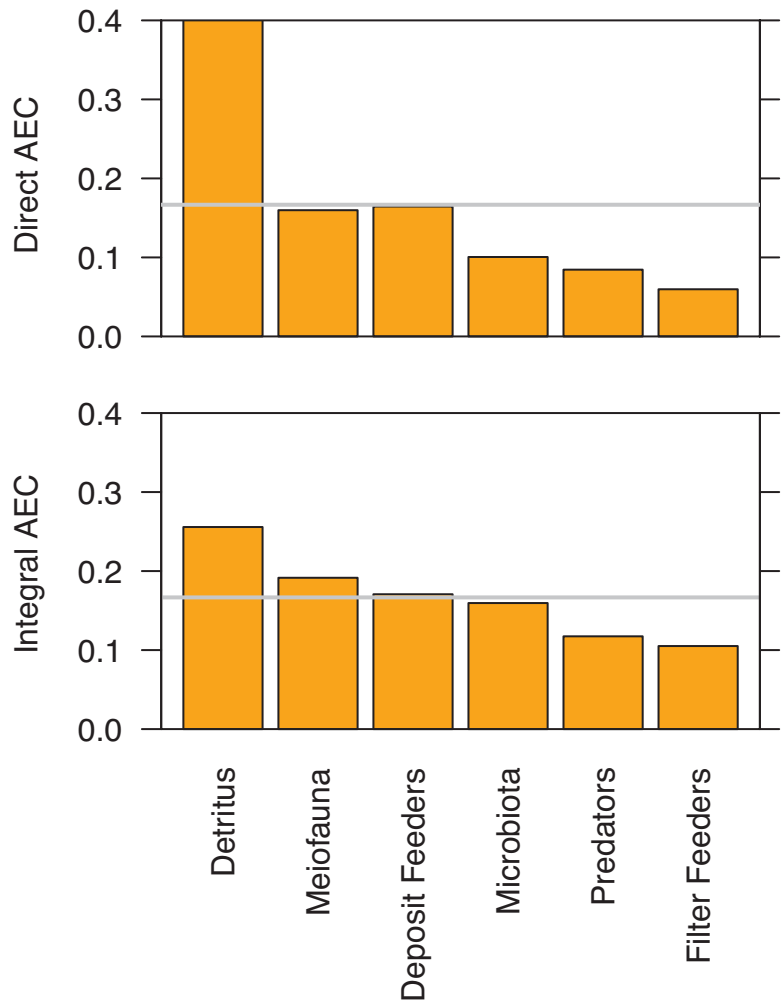}
  \caption{Comparing the direct and integral (direct, boundary,
    and indirect) $AEC$ distributions. The gray line represents the
    value if each node had the same centrality.}
	\label{fig:GNcomp}
\end{figure*}

\begin{figure*}[!t]
	\centering
   \includegraphics[scale=1]{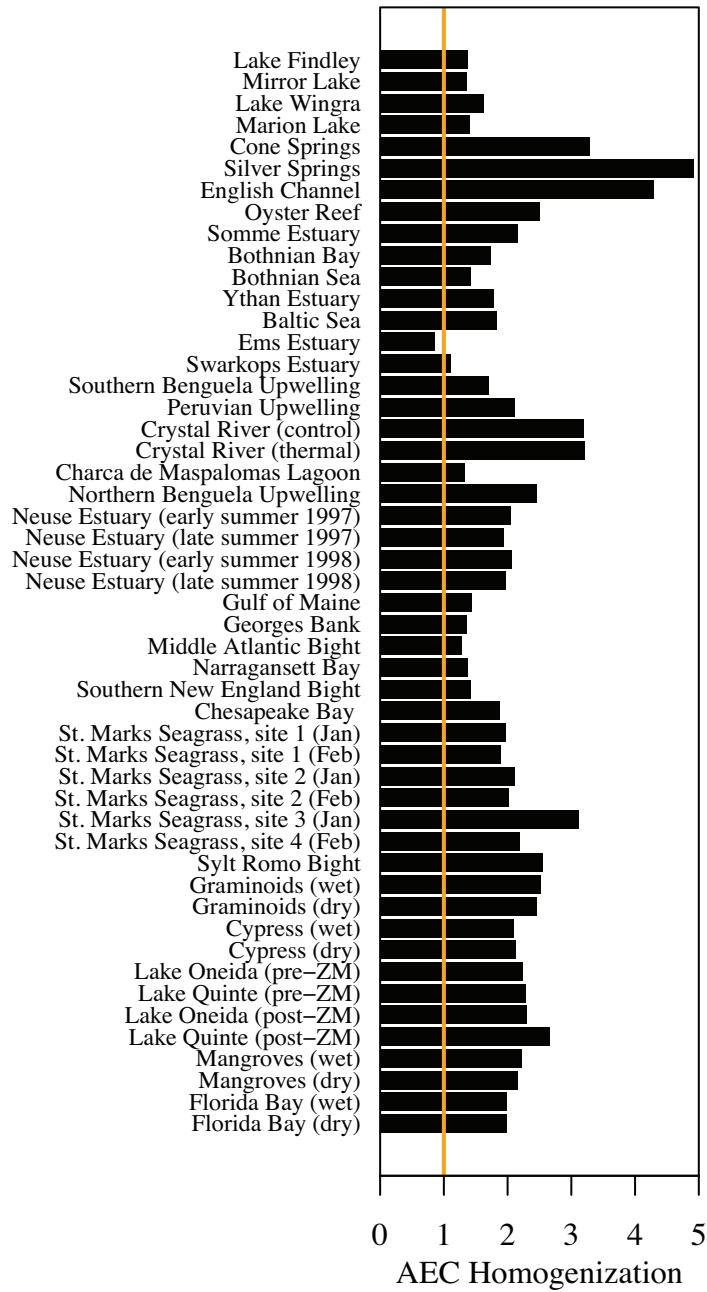}
   \caption{$CV$ ratios quantifying the homogenization of roles in
     $AEC$ when compared to $AEC^{direct}$. Bars surpassing the
     vertical line at 1 represent ecosystems which exhibit centrality
     homogenization from indirect effects.}
\label{fig:cv}
\end{figure*}

The comparison of direct and integral $AEC$ in the Oyster Reef model
(Fig.~\ref{fig:GNcomp}) illustrates how the relative importance of
species in generating $TST$ can become more similar when indirect
flows are considered.  Fig.~\ref{fig:cv} shows that this is a general
trend as 49 of the 50 network models (98\%) exhibited less variation
in $AEC$ when indirect flows were considered. The EMS estuary was the
only network to not meet our expectation. Furthermore, when boundary
effects were excluded from our analysis, 33 of 50 ecosystems (66\%)
exhibited homogenized importance due to indirect effects alone. Thus,
indirect effects appear to homogenize the importance of species in
ecosystems.

\section{Discussion}
 
\subsection{Ecological Significance}
Our results make two ecologically significant contributions.  First,
$AEC$ is a novel way to quantify the relative functional importance of
species or functional groups in an ecosystem with respect to the
global energy--matter exchange of a network. It is sensitive to both
extensive and intensive changes, and incorporates indirect effects.
With this metric, we gain new insight into the distribution of species
contribution to the total system activity.  Second, the results
highlight another critical consequence of indirect interactions in
ecosystems; they tend to make species' functional importance more
similar.  If this pattern holds generally, it has important
implications for conservation biology which we discuss below.
	 
Our results support the hypothesis that ecosystems are comprised of a
few functionally dominant species and a larger set of species with
relatively similar functional importance.  The smaller $CV$ values for the
remaining community after the dominants are excluded demonstrate that
the top ranked species are responsible for a majority of the variation
in $AEC$ and that most species in ecosystems have relatively low and
even centralities. This is further evidence for the general
dominance--diversity community patterns initially described by
\citet{whittaker65}.

The second hypothesis this paper tested concerned the effect of
indirect flows on the species' environ centralities. Our results
support the hypothesis that indirect effects tend to even the
functional importance of species.  Indirect flows alone homogenized the
importance of species in 66\% of the ecosystem networks analyzed. The
empirical literature is rich with examples of indirect relationships
driving ecosystem reactions \citep{berger08, bondavalli99,
  carpenter85, wootton94_review} and our theoretical analysis provides
evidence for another important consequence of indirect effects. 

The EMS estuary was the only network in which the integral centrality
$AEC$ was not more even than the direct environ centrality
$AEC^{Direct}$.  This could be because of the extreme dominant centrality
of the sediment POC in this model. After removing sediment POC from
the $AEC$ variation analysis, the $CV$ declined 70\% which was one of
the largest percent drops in variation. It was also one of the few
ecosystems to have a $CV(AEC)$ that was greater than unity. 

Collectively, our homogenization results support the trend towards
whole ecosystem management for conservation biology.  Although a few
species may have a dominant functional importance, most species are
contributing in similar ways.  This could be interpreted as suggesting
that losing the non-dominant species would have less significant
impacts on the system, but this misses a fundamental aspect of the
results.  The homogenization of the species centralities is a consequence of
the indirect interactions that are distributed across the network of
relationships.  This depends on the existing species and their pattern
of direct and indirect connections.  Thus these results show that
preserving the intact network is critical to maintain current
ecosystem total energy--matter throughflow.
		
\subsection{Centrality metric comparisons}
Our comparison of centrality metrics in Fig.~\ref{fig:comp}
highlights the differences and similarities between these five
metrics.  In the case of $WD$ and $WI$, both metrics are calculated
from the same matrix $\mathbf{F}$, but differ by the anchoring
metric (i.e. sum of energy--matter entering a node versus total
energy--matter flowing in the network) and the length of the pathways
considered. This may explain the similarities between $WD$ and $WI$ as
seen in the Oyster Reef model.

Average eigenvector centrality should be the most similar to average
environ centrality because they are both global centrality measures
related to pathways or flow.  However, our results show that these
centrality measures can be quite different.  The results of the
hypothetical ecosystem model are most instructive
(Fig.~\ref{fig:toy}).  The first and second model realizations whose
direct flow intensity distributions are the same have the same $AEVC$
distributions. The large drop in direct transfer efficiencies does not
affect the equilibrium distribution of flows, thus the eigenvector
centralities remain the same.  This is unlike $AEC$ which does respond
to the drop in transfer efficiency because it captures the transient
flow dynamics as well as the equilibrium distribution. This
demonstrates the ability of $AEC$ to capture extensive changes that
$AEVC$ does not. Both $AEC$ and $AEVC$ capture differences between
realizations three and four, demonstrating that both are able to
capture intensive changes to direct energy--matter flow. Although
$AEC$ and $AEVC$ are well correlated, most centrality metrics tend to
be well correlated \citep{valente08,newman06}, and small differences
between metrics can be biologically important (see Fig.~\ref{fig:comp2}).

Alternative centrality measures have different values.  We do not
expect one to be the universally correct analytical choice.  If the
goal is to understand the relative importance of nodes in generating
the total system activity of a network model of conservative flows, the
evidence presented here suggests that $AEC$ is a good choice.  

\subsection{Future work and limitations}
Despite the limitations of the set of ecosystem network models
(e.g. relatively few model authors, only 35 distinct ecosystems, some
very small models $n<5$, mass--balanced assumption), the consistency
of the results suggests they are robust to model error.  To verify the
generality of these results, this data set can be extended to include
biogeochemically-based networks that may reflect different aspects of
ecosystem organization \citep{borrett10_idd} as more models become
available. Our preliminary analysis suggest that these results may
hold true for biogeochemically--based networks, but this remains a
testable hypothesis.

One important direction for future research is clarifying the
relationship between alternative methods for quantifying
the ``importance'' and ``effects'' species have on one another
\citep{ulanowicz90, jordan03, schramski06, bauer10}.  For example,
\citeapos{schramski06} measure of distributed control among species in
a network is calculated using the input and output oriented integral
flow matrices.  There are some clear conceptual differences, but since
the underlying data and analytical goals are the same we expect there
to be an interesting relationship between $AEC$ and distributed
control.  Another possible comparison would be to calculate $WI$
beyond path lengths of 10 to see how longer indirect relationships
effect this centrality metric in an ENA framework.
 
\subsection{Summary}
This work makes three core contributions to ecology.  First, we
introduce a formal method called environ centrality to quantify the
relative importance of species in generating ecosystem activity. This
metric builds on the centrality concept from the social sciences and
Patten's environ concept, and can be considered in a manner similar to
rank-abundance curves familiar to many ecologists. The main ecological
advantages of environ centrality are that it is the only metric to
date to include all direct and indirect effects of a weighted
ecosystem network model, it is simple to calculate in the ENA
framework, and it should easily apply to network models of flow in
other types of complex systems.  Further, it is sensitive to both
extensive and intensive changes to direct flow intensities, and does
not ignore the sometimes important transient dynamics like eigenvector
centrality does.  Second, we applied environ centrality to find
evidence that support the hypothesis that ecosystems are comprised of
both a few functionally dominant species and a larger group of species
with more similar importance.  Our results expose the central
importance of detritus and decomposer species like bacteria and fungi
in generating carbon throughflow activity in ecosystems. This result
is not surprising, but it supports our claim that environ centrality
is a useful ecological measure. Third, this work shows that indirect
effects tend to homogenize the functional importance of species.  This
is further evidence of the important consequences that different types
of indirect effects can have in complex systems like ecosystems.

\section{Acknowledgments}
This research and manuscript benefited form critiques by Z.\ Long, A.\
Stapleton, M.\ Freeze, A.\ Salas, D.\ Hines, and M.\ Lau.  Two
anonymous reviewers also greatly improved the readability and
scholarship of this manuscript. We would also like to thank D. Baird,
R.R. Christian, J. Link, A.\ L.\ J.\ Miehls, U.\ Sharler, and R.\
E. Ulanowicz for sharing many of their ecosystem models with us.
Lastly, we gratefully acknowledge financial support from UNCW and NSF
DEB-1020944 (SRB), the NOAA Ernest F. Hollings Scholarship and the
UNCW Center for Support of Undergraduate Research and Fellowship
(SLF).



\newpage

\begin{table*}[h]		     
\centering
 \caption{Formulas and references for centrality indices compared in this paper}
\vspace{.1in}
\begin{tabular}{l l l l}
\hline
\hline
Metric	&Symbol	  &Formula &Citation\\
\hline
Degree Centrality&$DC$ &$(\sum_{i=1}^n a_{ij} + \sum_{j=1}^n
a_{ij})/2$ &\cite{freeman79}\\ [1ex]

Weighted Degree	&$WD$ &$(\sum_{i=1}^n f_{ij} + \sum_{j=1}^n f_{ij})/2$ &\cite{freeman79}\\[1.5ex]
Betweenness Centrality$^\dagger$	&$BC$ &$2\sum_{j<k;i\neq j}\frac{h_{jk}i/h_{jk}}{(n-1)(n-2)}$  $h_{jk}$ &\cite{freeman79}\\[1.5ex]
Weighted Importance     &$WI$ &$\sum_{k=1}^m \frac{\sum_{i=1}^n (\frac{f_{ij}}{\sum_{j=1}^n f_{ij}})^m}{m}$&\cite{jordan03,bauer10}\\[1.5ex]
Eigenvector Centrality$^\ddagger$     &$EVC$ &$(v + w)/2$&\cite{bonacich72}\\[1.5ex]
Environ Centrality      &$AEC$ &$(EC_{j}^{in} + EC_{j}^{out})/2$\\[1.5ex]	
\hline
\end{tabular}
\begin{footnotesize}

$^\dagger$ $h_{jk}i$ represents the number of shortest paths between node $j$ and $k$ that pass through node $i$.
$^\ddagger$ $v$ and $w$ are the right and left hand eigenvectors of the dominant eigen value respectively. 
\end{footnotesize}
	\label{tab:cent}
\end{table*}

\begin{table*}
  \caption{Fifty empirically-based trophic ecosystem network
    models.} \label{tab:models}
  \tableline 
\begin{center}
\begin{footnotesize}
\begin{tabular}{l l c c r c r}
Model & units & $n^\dagger$ & $C^\dagger$ & $TST^\dagger$ &
$FCI^\dagger$ & Source \\
\hline \\[-1.5ex]
Lake Findley & gC m$^{-2}$ yr$^{-1}$ & 4 & 0.38 & 51 & 0.30 & \citet{richey78} \\
Mirror Lake & gC m$^{-2}$ yr$^{-1}$ & 5 & 0.36 & 218 & 0.32 &  \citet{richey78} \\
Lake Wingra & gC m$^{-2}$ yr$^{-1}$ & 5 & 0.40 & 1,517 & 0.40 & \citet{richey78} \\
Marion Lake & gC m$^{-2}$ yr$^{-1}$ & 5 & 0.36 & 243 & 0.31 & \citet{richey78} \\
Cone Springs & kcal m$^{-2}$ yr$^{-1}$ & 5 & 0.32 & 30,626 & 0.09 & \citet{tilly68} \\
Silver Springs & kcal m$^{-2}$ yr$^{-1}$ & 5 & 0.28 & 29,175 & 0.00 & \citet{odum57} \\
English Channel & kcal m$^{-2}$ yr$^{-1}$ & 6 & 0.25 & 2,280 & 0.00 & \citet{brylinsky72} \\
Oyster Reef & kcal m$^{-2}$ yr$^{-1}$ & 6 & 0.33 & 84 & 0.11 & \citet{dame81} \\
Somme Estuary & mgC m$^{-2}$ d$^{-1}$ & 9 & 0.30 & 2,035 & 0.14 & \citet{rybarczyk03} \\
Bothnian Bay & gC m$^{-2}$ yr$^{-1}$ & 12 & 0.22 & 130 & 0.18 &  \citet{sandberg00} \\
Bothnian Sea & gC m$^{-2}$ yr$^{-1}$ & 12 & 0.24 & 458 & 0.27 &  \citet{sandberg00} \\
Ythan Estuary & gC m$^{-2}$ yr$^{-1}$ & 13 & 0.23 & 4,181 & 0.24 & \citet{baird81} \\
Baltic Sea & mgC m$^{-2}$ d$^{-1}$ & 15 & 0.17 & 1,974 & 0.13 &  \citet{baird91} \\
Ems Estuary & mgC m$^{-2}$ d$^{-1}$ & 15 & 0.19 & 1,019 & 0.32 & \citet{baird91} \\
Swarkops Estuary & mgC m$^{-2}$ d$^{-1}$ & 15 & 0.17 & 13,996 & 0.47 &  \citet{baird91} \\
Southern Benguela Upwelling & mgC m$^{-2}$ d$^{-1}$ & 16 & 0.23 & 1,774 & 0.19 &\citet{baird91} \\
Peruvian Upwelling & mgC m$^{-2}$ d$^{-1}$ & 16 & 0.22 & 33,496 & 0.04 & \citet{baird91} \\
Crystal River (control) & mgC m$^{-2}$ d$^{-1}$ & 21 & 0.19 & 15,063 & 0.07 & \citet{ulanowicz86} \\
Crystal River (thermal) & mgC m$^{-2}$ d$^{-1}$ & 21 & 0.14 & 12,032 & 0.09 & \citet{ulanowicz86} \\
Charca de Maspalomas Lagoon & mgC m$^{-2}$ d$^{-1}$ & 21 & 0.13 & 6,010,331 & 0.18 & \citet{almunia99} \\
Northern Benguela Upwelling & mgC m$^{-2}$ d$^{-1}$ & 24 & 0.21 & 6,608 &0.05 & \citet{heymans00} \\
Neuse Estuary (early summer 1997) & mgC m$^{-2}$ d$^{-1}$ & 30 & 0.09 & 13,826 & 0.12 & \citet{baird04} \\
Neuse Estuary (late summer 1997) & mgC m$^{-2}$ d$^{-1}$ & 30 & 0.11 & 13,038 & 0.13 & \citet{baird04} \\
Neuse Estuary (early summer 1998) & mgC m$^{-2}$ d$^{-1}$ & 30 & 0.09 & 14,025 & 0.12 & \citet{baird04} \\
Neuse Estuary (late summer 1998) & mgC m$^{-2}$ d$^{-1}$ & 30 & 0.10 & 15,031 & 0.11 & \citet{baird04} \\
Gulf of Maine & g ww m$^{-2}$ yr$^{-1}$ & 31 & 0.35 & 18,382 & 0.15 &  \citet{link08} \\
Georges Bank & g ww m$^{-2}$ yr$^{-1}$ & 31 & 0.35 & 16,890 & 0.18 & \citet{link08} \\
Middle Atlantic Bight & g ww m$^{-2}$ yr$^{-1}$ & 32 & 0.37 & 17,917 & 0.18 & \citet{link08} \\
Narragansett Bay & mgC m$^{-2}$ yr$^{-1}$ & 32 & 0.15 & 3,917,246 & 0.51  & \citet{monaco97} \\
Southern New England Bight & g ww m$^{-2}$ yr$^{-1}$ & 33 & 0.03 & 17,597 & 0.16 & \citet{link08} \\
Chesapeake Bay  & mgC m$^{-2}$ yr$^{-1}$ & 36 & 0.09 & 3,227,453 & 0.19 & \citet{baird89} \\
St. Marks Seagrass, site 1 (Jan) & mgC m$^{-2}$ d$^{-1}$ & 51 & 0.08 & 1,316 & 0.13 & \citet{baird98} \\
St. Marks Seagrass, site 1 (Feb) & mgC m$^{-2}$ d$^{-1}$ & 51 & 0.08 & 1,591 & 0.11 & \citet{baird98} \\
St. Marks Seagrass, site 2 (Jan) & mgC m$^{-2}$ d$^{-1}$ & 51 & 0.07 & 1,383 & 0.09 & \citet{baird98} \\
St. Marks Seagrass, site 2 (Feb) & mgC m$^{-2}$ d$^{-1}$ & 51 & 0.08 & 1,921 & 0.08 & \citet{baird98} \\
St. Marks Seagrass, site 3 (Jan) & mgC m$^{-2}$ d$^{-1}$ & 51 & 0.05 & 12,651 & 0.01 &\citet{baird98} \\
St. Marks Seagrass, site 4 (Feb) & mgC m$^{-2}$ d$^{-1}$ & 51 & 0.08 & 2,865 & 0.04 & \citet{baird98} \\
Sylt-R{\o}m{\o} Bight & mgC m$^{-2}$ d$^{-1}$ & 59 & 0.08 & 1,353,406 & 0.09 & \citet{baird04_sylt} \\
Graminoids (wet) & gC m$^{-2}$ yr$^{-1}$ & 66 & 0.18 & 13,677 & 0.02 & \citet{ulanowicz00_graminoids} \\
Graminoids (dry) & gC m$^{-2}$ yr$^{-1}$ & 66 & 0.18 & 7,520 & 0.04 &  \citet{ulanowicz00_graminoids} \\
Cypress (wet) & gC m$^{-2}$ yr$^{-1}$ & 68 & 0.12 & 2,572 & 0.04 & \citet{ulanowicz97_cypress} \\
Cypress (dry) & gC m$^{-2}$ yr$^{-1}$ & 68 & 0.12 & 1,918 & 0.04 & \citet{ulanowicz97_cypress}\\
Lake Oneida (pre-ZM) & gC m$^{-2}$ yr$^{-1}$ & 74 & 0.22 & 1,638 & $<0.01$ & \citet{miehls09_oneida} \\
Lake Quinte (pre-ZM) & gC m$^{-2}$ yr$^{-1}$ & 74 & 0.21 & 1,467 & $<0.01$ &  \citet{miehls09_quinte} \\
Lake Oneida (post-ZM) & gC m$^{-2}$ yr$^{-1}$ & 76 & 0.22 & 1,365 & $<0.01$ & \citet{miehls09_oneida} \\
Lake Quinte (post-ZM) & gC m$^{-2}$ yr$^{-1}$ & 80 & 0.21 & 1,925 & $0.01$ &  \citet{miehls09_quinte} \\
Mangroves (wet) & gC m$^{-2}$ yr$^{-1}$ & 94 & 0.15 & 3,272 & 0.10 & \citet{ulanowicz99_mangrove} \\
Mangroves (dry) & gC m$^{-2}$ yr$^{-1}$ & 94 & 0.15 & 3,266 & 0.10 & \citet{ulanowicz99_mangrove} \\
Florida Bay (wet) & mgC m$^{-2}$ yr$^{-1}$ & 125 & 0.12 & 2,721 & 0.14 & \citet{ulanowicz98_fb} \\
Florida Bay (dry) & mgC m$^{-2}$ yr$^{-1}$ & 125 & 0.13 & 1,779 & 0.08& \citet{ulanowicz98_fb} \\[0.5ex] 
\end{tabular}
\end{footnotesize}
\end{center}
\tableline
\begin{scriptsize}
$^\dagger$ $n$ is the number of nodes in the network model, $C=L/n^2$
is the model connectance when $L$ is the number of direct links or
energy--matter transfers, $TST=\sum\sum{f_{ij}}+\sum{z_i}$ is the total
system throughflow, and $FCI$ is the Finn Cycling Index \citep{finn80}.  
\end{scriptsize}
\end{table*}

\newpage


\begin{table*}
\center
\caption{Dominate species and functional variation in 50 ecosystem network models}\label{tab:dom}
\begin{scriptsize}
\begin{tabular}{l r r c l }
\hline
\hline
Model & CV(AEC) & CV(R-AEC)$^\dagger$ & \# Dominants & Dominant Identity \\
\hline
Lake Findley  & 0.48 & 0.48 & 0 \\
Mirror Lake  & 0.51 & 0.17 & 2 & Sediment, Benthos \\
Lake Wingra  & 0.43 & 0.43 & 0 \\
Marion Lake  & 0.53 & 0.14 & 2 & Sediment, Benthos \\
Cone Springs  & 0.23 & 0.23 & 0 \\
Silver Springs  & 0.11 & 0.11 & 0 \\
English Channel  & 0.16 & 0.16 & 0 \\
Oyster Reef  & 0.33 & 0.33 & 0 \\
Somme Estuary  & 0.58 & 0.18 & 1 & POM \\
Bothnian Bay  & 0.34 & 0.34 & 0 \\
Bothnian Sea  & 0.46 & 0.46 & 0 \\
Ythan Estuary  & 0.50 & 0.50 & 0 \\
Baltic Sea  & 0.40 & 0.40 & 0 \\
Ems Estuary  & 1.03 & 0.31 & 1 & Sediment POC \\
Swarkops Estuary  & 0.97 & 0.30 & 1 & Sediment POC \\
Southern Benguela Upwelling  & 0.58 & 0.26 & 1 & Suspended POC \\
Peruvian Upwelling  & 0.41 & 0.41 & 0 \\
Crystal River (control)  & 0.56 & 0.27 & 1 & Detritus \\
Crystal River (thermal)  & 0.61 & 0.30 & 1 & Detritus \\
Charca de Maspalomas Lagoon  & 0.70 & 0.40 & 1 & Sedimented POC \\
Northern Benguela Upwelling  & 0.58 & 0.23 & 1 & POC \\
Neuse Estuary (early summer 1997)  & 0.79 & 0.43 & 2 & Sediment POC, Sediment bacteria \\
Neuse Estuary (late summer 1997)  & 0.83 & 0.37 & 2 & Sediment POC, Sediment bacteria \\
Neuse Estuary (early summer 1998)  & 0.80 & 0.44 & 2 & Sediment POC, Sediment bacteria \\
Neuse Estuary (late summer 1998)  & 0.81 & 0.41 & 2 & Sediment POC, Sediment bacteria \\
Gulf of Maine  & 0.65 & 0.41 & 2 & Detritus/POC, Other Macrobenthos \\
Georges Bank  & 0.68 & 0.41 & 2 & Detritus/POC, Other Macrobenthos \\
Middle Atlantic Bight  & 0.73 & 0.45 & 2 & Detritus/POC, Other Macrobenthos \\
Narragansett Bay  & 1.34 & 0.39 & 3 & \parbox[t][5.5ex]{5.25 cm}{Detritus/POC, Sediment POC Bacteria, Pelagic Bacteria} \\
Southern New England Bight  & 0.68 & 0.49 & 1 & Detritus/POC \\
Chesapeake Bay   & 1.10 & 0.36 & 2 & \parbox[t][5.5ex]{5.25cm}{Sedimented POC, Bacteria in Sediment (POC)} \\
St. Marks Seagrass, site 1 (Jan.)  & 1.00 & 0.38 & 2 & Sediment POC, Benthic Bacteria \\
St. Marks Seagrass, site 1 (Feb.)  & 0.94 & 0.42 & 2 & Sediment POC, Benthic Bacteria \\
St. Marks Seagrass, site 2 (Jan.)  & 0.86 & 0.34 & 2 & Sediment POC, Benthic Bacteria \\
St. Marks Seagrass, site 2 (Feb.)  & 0.86 & 0.35 & 2 & Sediment POC, Benthic Bacteria \\
St. Marks Seagrass, site 3 (Jan.)  & 0.60 & 0.27 & 1 & Sediment POC \\
St. Marks Seagrass, site 4 (Feb.)  & 0.75 & 0.34 & 1 & Sediment POC \\
Sylt-R{\o}m{\o} Bight  & 1.17 & 0.44 & 1 & Sediment POC \\
Graminoids (wet)  & 0.98 & 0.18 & 2 & Sediment Carbon, Refractory Detritus \\
Graminoids (dry)  & 1.00 & 0.28 & 2 & Sediment Carbon, Refractory Detritus \\
Cypress (wet)  & 0.89 & 0.49 & 3 & \parbox[t][5.5ex]{5.25 cm}{Liable Detritus, Living Sediment, Vertebrate Detritus} \\
Cypress (dry)  & 0.85 & 0.50 & 3 & \parbox[t][5.5ex]{5.25 cm}{Vertebrate Detritus, Liable Detritus, Living Sediment}  \\
Lake Oneida (pre-ZM)  & 0.49 & 0.49 & 0 \\
Lake Quinte (pre-ZM)  & 0.62 & 0.28 & 1 & Sedimented Detritus \\
Lake Oneida (post-ZM)  & 0.49 & 0.49 & 0 \\
Lake Quinte (post-ZM)  & 0.78 & 0.29 & 1 & Sedimented Detritus \\
Mangroves (wet)  & 1.17 & 0.35 & 3 & \parbox[t][5.5ex]{5.25 cm}{Sediment Carbon, Sediment bacteria, POC} \\
Mangroves (dry)  & 1.19 & 0.35 & 3 & \parbox[t][5.5ex]{5.25 cm}{Sediment Carbon, Sediment bacteria, POC} \\
Florida Bay (wet)  & 1.43 & 0.46 & 3 & \parbox[t][5.5ex]{5.25 cm}{Water POC, Benthic POC, Water Flagellates} \\
Florida Bay (dry)  & 1.40 & 0.36 & 3 & \parbox[t][5.5ex]{5.25cm}{Water POC, Benthic POC, Water Flagellates} \\
\end{tabular}
\end{scriptsize}
\tableline

\begin{footnotesize}
$^\dagger$ CV(R-AEC) represents the $CV$ of $AEC$ scores for all
  species remaining after dominates have been removed.  
\end{footnotesize}

\end{table*}


\clearpage

\end{spacing} 
\end{document}